\documentclass{aa}
\usepackage{graphicx}
\begin{document}


\title{The complementarity of astrometric and radial velocity exoplanet observations}
\subtitle{Determining exoplanet mass with astrometric snapshots}

\author{Mikko Tuomi\inst{1,}\inst{2}\thanks{\email{mikko.tuomi@utu.fi}} \and Samuli Kotiranta\inst{2}\thanks{\email{samuli.kotiranta@utu.fi}} \and Mikko Kaasalainen\inst{1}}

\institute{Department of Mathematics and Statistics, P.O.Box 68, FI-00014 University of Helsinki, Finland \and Tuorla Observatory, V\"ais\"al\"antie 20, FI-21500, Piikki\"o, University of Turku, Finland}

\date{Received 30.5.2008 / Accepted 3.11.2008}

\abstract{}
{It is commonly assumed that the two indirect exoplanet detection methods, the radial velocity method and astrometric method, require observational periods exceeding the orbital period to produce positive results. Here we test this assumption in detail. We also investigate the smallest ratio of observational timeline and orbital period required for positive detections.}
{We obtain full information on the orbital parameters by combining radial-velocity and astrometric measurements by means of Bayesian inference, and sample the parameter probability densities of orbital and other model parameters with a Markov Chain Monte Carlo (MCMC) method in simulated observational scenarios to test the detectability of planets with orbital periods longer than the observational timelines.}
{We show that, when fitting model parameters simultaneously to measurements from both sources, it is possible to extract much more information from the measurements than when using either source alone. Currently available high-precision measurements of radial velocity (with 1ms$^{-1}$ precision) and astrometric measurements achievable with the SIM space telescope (with a precision of 1$\mu$as) can be used together to detect a Jupiter analog 30pc away with an observational timeline of only three years, approximately one fourth of the orbital period. Such measurements are sufficient for determining all its orbital parameters, including inclination and the true mass. Also, with accurate radial velocity measurements covering a timeline of 20 years, the true mass could be determined by astrometric observations within a single year. These case studies demonstrate the potential power of the Bayesian inference of multiple data sources in exoplanet observations. As an example, we show that using the currently available radial velocity measurements, the inclination of HD 154345b could be determined with SIM in a year.}
{}


\titlerunning{Astrometric snapshots}

\authorrunning{M. Tuomi et al.}

\maketitle

\section{Introduction}

Since the discovery of 51 Peg b (Mayor \& Queloz 1995), high-precision radial velocity (RV) measurements have been used successfully to detect planetary companions of nearby stars (Butler et al. 2006). The finest instruments to date are capable of achieving an RV precision of 1ms$^{-1}$ (Santos et al. 2004; Moorwood and Masanori, 2004). However, the exact nature of the companions remains unknown when only RV measurements are available, as these only yield the product of mass and sine of the orbital inclination, giving the lower limit for the mass. Recently, it has been claimed that an RV precision of 1cms$^{-1}$ could be technically possible in the future (Li et al. 2008). However, the RV variations of stars of approximately 5ms$^{-1}$ to 50ms$^{-1}$ for K to F stars, caused by star spots and irregular convective zones, will likely prevent the detection of signals of the cm scale (Saar \& Donahue 1997; Saar et al. 1998).

Astrometric (A) measurements of the position of the target star in the sky as a function of time can be used to detect the inclination, and consequently the true mass of the companion. But despite several trials (van de Kamp 1969; Neuhaeuser et al. 2008), the precision of these measurements has not been high enough to verify the planetary nature of RV companions.

With the aid of telescopes and instruments capable of high-precision astrometry (SIM, GAIA, PRIMA, etc.); however, this situation is about to change. With an estimated astrometric precision of future telescopes of 1.0$\mu$as (Unwin et al. 2008) or 8-10$\mu$as (Casertano et al. 2008; Derie et al. 2002), it will be possible to determine the masses of the already detected RV companions. It is commonly assumed that these detections require observational periods longer than the orbital period of the target system to be able to detect the periodic signal. Here we test this assumption in detail.

In this article we simulate astrometric and RV measurements to study the possibility of detecting planetary companions of nearby stars with various observational timelines. The goal is to find the minimum timeline required for detecting a planetary companion using high-precision RV and astrometric measurements. These two sources of data are combined by means of Bayesian inference, and the probability densities of orbital and inertial reference frame parameters are sampled using Markov Chain Monte Carlo (MCMC) (Metropolis et al. 1953; Hastings 1970) to find the full global solution to this multidata inverse problem (see Ford 2005; Ford 2006; Maness et al. 2007). Also, the probability densities are sampled to calculate the realistic error bars for parameters and to determine whether a positive detection has been made or not. We also analyse the RV measurements of HD 154345 (Wright et al. 2008) alone and together with simulated astrometric measurements to estimate the minimum observational timeline of astrometry for the detection of the true mass of HD 154345 b.


\section{Modelling the data}

The motion of a planet around a star was treated as a simple two-body system, with masses $m_{\star}$ and $m_{p}$ for the star and the planetary companion, respectively. In Cartesian coordinates, when the gravitational forces between the possible other planets in the system are assumed negligible, the column position vector of the star with respect to the barycentre of the system can be expressed as a function of time ($t$) as (e.g. Green 1985)
\begin{equation}\label{model}
\begin{array}{lll}
    \vec{R}(t) & = & \vec{R}(0) + \dot{\vec{R}}(0)t + \vec{P} [\cos E(t) -e] \\
    & & + \vec{Q} \sqrt{1-e^{2}} \sin E(t) ,
\end{array}
\end{equation}
where $E$ is the eccentric anomaly, satisfying the Kepler equation $E - e \sin E = 2 \pi n (t-t_{0})$, and $n = P^{-1}$ is the orbital frequency, $P$ the orbital period and $t_{0}$ the time of periastron. The velocity and position $\dot{\vec{R}}(0)$ and $\vec{R}(0)$ w.r.t. the observer are some constant vectors defining the inertial reference frame. Vectors $\vec{P}$ and $\vec{Q}$ are constant vectors, defined as
\begin{equation}\label{thieleinnes}
 \left\{\begin{array}{ll}
    \vec{P} = a_{\star} ( \vec{l} \cos \omega + \vec{k} \sin \omega ) \\
    \vec{Q} = a_{\star} ( - \vec{l} \sin \omega + \vec{k} \cos \omega ) \\
    \vec{l} = (\sin \Omega, \cos \Omega, 0) \\
    \vec{k} = (I \cos \Omega, - I \sin \Omega, \sin i) \\
\end{array}\right. .
\end{equation}
The parameters $a_{\star}$, $e$, $\Omega$, $I = \cos i$, and $\omega$ are the orbital parameters of the system: semimajor axis of the star, eccentricity, longitude of ascending node, cosine of the inclination and the longitude of pericentre, respectively. The semimajor axis of the star can be expressed as a function of the masses of the gravitationally interacting bodies,
\begin{equation}\label{massfunction}
    a_{\star} = m_{p} G^{1/3} (n [m_{\star} + m_{p}])^{-2/3} .
\end{equation}

The models for astrometric and RV data are now simply $\vec{\Theta}(t) = [\Theta_{x}(t), \Theta_{y}(t)] = D^{-1} [R_{x}(t), R_{y}(t)]$ for astrometry and $\dot{z}(t) = \partial_{t} R_{z}(t)$ for RV. Here $D$ is the distance of the system from the observer. There are now 12 independent parameters describing the system. The parameter vector $\vec{u}$ of parameter space $U$ can be written as $\vec{u} = (\lambda_{x}, \lambda_{y}, \mu_{x}, \mu_{y}, \gamma, a_{\star}, \omega, I, \Omega, e, t_{0}, n) \in U$, where $\lambda_{x} = D^{-1} \dot{R}_{x}(0)$, $\lambda_{y} = D^{-1} \dot{R}_{y}(0)$, $\mu_{x} = D^{-1} R_{x}(0)$, $\mu_{y} = D^{-1} R_{y}(0)$ and $\gamma = \dot{R}_{z}(0)$. To fully describe the system, the masses $m_{p}$ and $m_{\star}$ should be treated as independent unknown parameters, and the distance $D$ should also be included in vector $\vec{u}$. Here we assume that $m_{\star}$ and $D$ are known with sufficient accuracy by some astrophysical techniques.

It is essential to include the parameters defining the inertial reference frame to be able to fully investigate the probability densities of the parameters. It is also assumed that the model parameters do not change as a function of time during the observational timeline.

It is intuitively clear that the availability of both RV and A data sets, instead of only one of them, should increase the amount and the quality of information on the observed system. This should be true even if one of the data sets is significantly more inaccurate than the other. In what follows, we state this principle more rigorously and present a practical method for combining the two data sources. In fact, since RV and A data are radial and tangential projection samples of orbital motion, they are separate sources only in the instrumental sense, so we could just as easily talk about full velocity data.

\section{Simulations and fitting procedure}

We simulated data sets to study the complementarity of astrometric and RV time series. Assuming that the errors $\epsilon_{i}$ for each data point $i = 1,...,N$ are independent and identically distributed and that their probability density functions (PDF's) are Gaussian ($\epsilon_{i} \sim N(0,\sigma^{2}_{i})$), the observed RV data are modelled as
\begin{equation}\label{rvmodel}
    \dot{z}_{i} = \dot{z}(\vec{u},t_{i}) + \epsilon_{i} .
\end{equation}
Equally, the two-dimensional astrometric model can be written as
\begin{equation}\label{asmodel}
    \vec{\Theta}_{i} = \vec{\Theta}(\vec{u},t_{i}) +
    \vec{\epsilon}_{i} .
\end{equation}
To avoid inversion crimes, i.e. to avoid using exactly the same model to generate the measurements and to find the inverse solution (see e.g. Kaipio and Somersalo 2005), an additional planet of low mass was included in the model when generating the data. This choice was made because small systematic errors make the simulated measurements more realistic. Otherwise the simulated measurements and the corresponding solutions would only help for studying the model in Eq. (\ref{model}), not necessarily situations encountered in reality.

Several data sets were generated, each with a different value for the data parameter $T$, representing the length of the observational timeline. It was further assumed that the observation time $t_{i}$ was evenly distributed in $T$.

When simulating the measurements, the timelines were set to $T_{A} = T_{RV} =$ 20.0, 10.0, 5.0, 3.0, 1.5, 1.2, 1.0 and 0.8 years. The values of all the other data and orbital parameters and masses were fixed in all these scenarios. These values were set to $(\sigma_{RV}, \sigma_{A}, N_{RV}, N_{A}) = $ (1.0m/s, 1.0$\mu$as, 100, 100) and $(a, e, \omega, i, \Omega, t_{0}, m_{p}, m_{\star}, D) = $ (5.0AU, 0.1, 1.0, 1.0, 1.0, 1000.0d, M$_{J}$, M$_{\odot}$, 30pc), where $M_{J}$ is the mass of Jupiter. This simulated system is called S1.

\subsection{What is a positive detection?}

In the simplest possible case, when $e = 0$ and $i = \pi/2$, the detection threshold of full velocity data can be calculated analytically. Let $N$ be the number of data points, $T$ the length of the timeline of observations, and $\sigma$ the standard deviation of observations. Following the approach in Eisner \& Kulkarni (2001a, 2001b), the detected signal of the velocity variation amplitude is a false one produced by the uncertainties in the data with a probability $<1$\% if
\begin{eqnarray}\label{amplitude}
  a_{\star}^{2} & > & 4.61 \Bigg[ \Big( \frac{N_{RV}}{2 \sigma^{2}_{c,RV}} + \frac{N_{A}}{2 \sigma^{2}_{c,A}} \Big)^{-1} \nonumber\\
  & & + \Big( \frac{N_{RV}}{2 \sigma^{2}_{s,RV}} + \frac{N_{A}}{2 \sigma^{2}_{s,A}} \Big)^{-1} \Bigg]
\end{eqnarray}
where
\begin{eqnarray}\label{vardef}
\sigma_{c,RV} & = & \left\{ \begin{array}{ll}
  P \sigma_{RV} & , \frac{T_{RV}}{P} \geq 1 \\
  2 P \sigma_{RV} \Big[ 1- \cos \big( \frac{\pi T_{RV}}{P} \big) \Big]^{-1} & , \frac{T_{RV}}{P} < 1 \\
\end{array} \right. \\
\sigma_{c,A} & = & \left\{ \begin{array}{ll}
  D \sigma_{A} & , \frac{T_{A}}{P} \geq 1 \\
  2 D \sigma_{A} \Big[ 1- \cos \big( \frac{\pi T_{A}}{P} \big) \Big]^{-1} & , \frac{T_{A}}{P} < 1 \\
\end{array} \right.  \\
\sigma_{s,RV} & = & \left\{ \begin{array}{ll}
  P \sigma_{RV} & , \frac{T_{RV}}{P} \geq \frac{1}{2} \\
  P \sigma_{RV} \Big[ \sin \big( \frac{\pi T_{RV}}{P} \big) \Big]^{-1} & , \frac{T_{RV}}{P} < \frac{1}{2} \\
\end{array} \right.  \\
\sigma_{s,A} & = & \left\{ \begin{array}{ll}
  D \sigma_{A} & , \frac{T_{A}}{P} \geq 2 \\
  2 \pi D \sigma_{A} \Big[\frac{\pi T_{A}}{P} - \sin \big( \frac{\pi T_{A}}{P} \big) \Big]^{-1} & , \frac{T_{A}}{P} < 2 \\
\end{array} \right.
\end{eqnarray}
and $\sigma_{A}$ is in radians. This approach excludes the uncertainties in the orbital period and can therefore only yield the lower limit for the detection threshold. Hence, if Eq. (\ref{amplitude}) does not hold, it will be impossible to detect the signal. However, if it holds, the detectability of such a companion needs to be examined more closely by numerical simulations and by analysing the simulated data using methods such as MCMC and Bayesian model selection criterion.

To fully investigate the ability to detect planetary companions, we must define when a positive detection has been made. This question can be approached through Bayesian probabilities. Let $\vec{R}_{1}$ be the model in Eq. (\ref{model}) with one planetary companion (corresponding 12 parameters in the RV and astrometry models), and $\vec{R}_{0}$ a model without a planetary companion (5 parameters). In general, let $\vec{R}_{k}$ be a model with $k$ planets.

Using the Bayes theorem, it can be seen that the conditional probability of model $\vec{R}_{k}$ representing the data ($m$) best, out of the $p+1$ alternatives to be tested, can be written as
\begin{equation}\label{model_probability}
  P(\vec{R}_{k} | m) = P(\vec{R}_{k}) \Bigg[ \sum_{j=0}^{p} B_{k,j}(m) P(\vec{R}_{j}) \Bigg]^{-1} ,
\end{equation} 
where the Bayes factor $B_{k,j}$ is defined as (e.g. Kass and Raftery, 1995)
\begin{equation}\label{bayes_factor}
  B_{k,j}(m) = \frac{P(m | \vec{R}_{k})}{P(m | \vec{R}_{j})},
\end{equation}
and $P(\vec{R}_{k})$ is the prior probability of the $k$th model, here set equal for all $k$, because it is assumed that there is no prior information available. Here the likelihood $P(m | \vec{R}_{k})$, with parameters $\vec{u}_{k} \in U_{k}$ for the $k$th model, is
\begin{equation}\label{parameter_integral}
  P(m | \vec{R}_{k}) = \int_{\vec{u}_{k} \in U_{k}} p(m | \vec{u}_{k}, \vec{R}_{k}) p(\vec{u}_{k} | \vec{R}_{k}) d \vec{u}_{j} ,
\end{equation}
where $p(m | \vec{u}_{k}, \vec{R}_{k})$ is the parameter likelihood function and $p(\vec{u}_{k} | \vec{R}_{k})$ the prior density.

Since the model probability, defined in this way, automatically takes the Occamian principle of parsimony into account, the model with the smallest number of parameters out of those having almost equal probabilities will be selected. Hence, it can be said that a detection has been made if (Jeffreys 1961)
\begin{equation}\label{best_model}
  P(\vec{R}_{1} | m) \gg P(\vec{R}_{0} | m) .
\end{equation}
This criterion is used throughout this article when deciding whether a statistically significant detection has been made or not.

\subsection{Fitting method}

The fitting was performed by requiring that the values of all the least-squares cost-functions $S_{x}$ (astrometric $x$), $S_{y}$ (astrometric $y$), $S_{RV}$ (RV), and their sum be minimized simultaneously. This method, called multidata inversion, has been used successfully with astrometric and RV measurements when detecting stellar binaries (e.g. Torres 2007). See the discussion in Kaasalainen and Lamberg (2006), where the multidata inversion was applied to asteroid observations.

The models for astrometric position and RV of the two-body system of interest are non-linear, so an iterative method of fitting the model parameters is needed. The MCMC with Metropolis-Hastings (M-H) algorithm was chosen because it is a global method (Metropolis et al. 1953; Hastings 1970), it offers a direct estimate of the posterior probability density, and because it can be used to verify the existence and uniqueness of the solution. Since the probability densities given the measurements are available, MCMC can be used to calculate realistical error estimates for the model parameters. These estimates are typically much larger than those calculated using traditional methods (e.g. Ford 2006), implying that MCMC should be preferred when assessing the parameter errors. Assuming Gaussian errors with zero mean, the likelihood function of the parameters with respect to RV measurements can be written as
\begin{equation}\label{likelihood}
  p(\dot{\vec{z}} | \vec{u}) \propto \exp ( -0.5 S_{RV} ) .
\end{equation}
When applying MCMC, a parameter value ($\vec{u}_{0}$) is selected for the first member of the chain. The next value $\vec{u}_{k+1}$ is found by randomly selecting a proposal in the vicinity of $\vec{u}_{k}$. This is then accepted by comparing the likelihoods of the two parameter values. Proposed parameter values $\vec{u}_{k+1}$ with a greater likelihood than that of $\vec{u}_{k}$ are always selected as the next chain member, but values with a smaller likelihood can also be selected according to the criterion of Hastings (1970). Samples of at least $10^{5}$ points were generated when sampling the parameter space. For practical details on MCMC with astronomical data, see e.g. Gregory (2005).

The parameter space $U$ in this Keplerian two-body model has a comparatively small dimension ($\dim U = 12$), but in some cases it already makes the sampling computationally expensive. Especially when covariances between the parameters are large and of non-linear nature, the space of reasonable probability $U_{R} \subset U$ to be sampled can be very narrow and, as a result, the next proposed value of parameter vector $\vec{u}$ in the Markov chain is likely to be outside this subspace and thus rejected, considerably increasing the time needed to generate a statistically representative chain. For this reason, when using a multivariate Gaussian density as a proposal, the acceptance rates were low, approximately 0.1 in the MCMC samplings.

\section{Taking advantage of Bayesian inference}

With more than one source of measurements available, it is possible to get more information from the system of interest than when relying on any single observation method alone. This is a consequence of Bayesian inference.

Denoting the astrometric measurements by $\vec{\Theta}_{o}$ and the RV measurements by $\dot{\vec{z}}_{o}$, the conditional probability of having parameter vector $\vec{u} \in U$, is a product of the impacts the two sets of measurements have on this hypothesis:
\begin{equation}\label{model probability}
  p(\vec{u} | \dot{\vec{z}}_{o},\vec{\Theta}_{o}) = \frac{p(\dot{\vec{z}}_{o} | \vec{u})}{p(\dot{\vec{z}}_{o})} \frac{p(\vec{\Theta}_{o} | \vec{u})}{p(\vec{\Theta}_{o})} p(\vec{u}) .
\end{equation}
Thus, there is always more information available on the system -- either in a narrower parameter density or in the possibility of including more parameters in the model -- when using multiple data sources. This is due to complementary rather than just additional information: the separate probability densities for complementary sources are quite different from each other, so their product (joint probability) is much more tightly bound than either factor alone.

Equation (\ref{model probability}) is in fact just another way of stating that we simply minimized the sum $S = S_{x} + S_{y} + S_{RV}$, as can be seen by applying the Eq. (\ref{likelihood}), while checking that each $S_{i}$ was still close to their minima. However, when calculating model probabilities and parameter densities, the formulation in Eq. (\ref{model probability}) has to be used.

\subsection{Correlations and complementarity}

By having measurements made using two different observational techniques has an effect on the parameter PDF's. This happens because the two measurements are modelled using a different model with differing parameters for the inertial reference frame. Therefore, it is expected that the two measurements contain complementary information on different aspects of the system.

Generally, correlations between model parameters occur if, for some small displacement of the parameter vector, $\delta \vec{u}$, the model $\vec{R}$ used to describe the system satisfies
\begin{equation}\label{unchanged}
    \vec{R}(\vec{u}) - \vec{R}(\vec{u} + \delta \vec{u}) \approx 0 .
\end{equation}
Now the parameter PDF's are broadened or correlated until Eq. (\ref{unchanged}) no longer holds.

In a stellar system with a single planetary companion, the most obvious possible coupling, a positive correlation between $t_{0}$ and $\omega$, is a natural byproduct of the two-body Keplerian model. When assuming $e \approx 0$ and using Eq. (\ref{thieleinnes}), Eq. (\ref{model}) becomes
\begin{equation}\label{e0further}
\begin{array}{lll}
    \vec{R}(t) & \approx & m [ \vec{l} \cos (\omega +n t - n t_{0}) \\
    & & + \vec{k} \sin (\omega + n t - n t_{0})] + \dot{\vec{R}}(0)t +\vec{R}(0) .
\end{array}
\end{equation}
Setting $\vec{R}(\omega, t_{0}) - \vec{R}(\omega + \delta \omega, t_{0}+\delta t_{0}) = 0$ implies $\delta \omega = n \delta t_{0}$, resulting in a positive linear correlation between parameters $\omega$ and $t_{0}$.

If it is also assumed that the observational timeline is much shorter than the orbital period, $T \ll P$, more correlations take place in this long-period system. As $t_{i} \in [-T/2, T/2]$ for all $i = 1, ..., N$, when $T/P \rightarrow 0$, each $t_{i}/P \rightarrow 0$ as well. Thus, this assumption justifies $\cos(n t) \approx 1$ and $\sin(n t) \approx n t$ and Eqs. (\ref{rvmodel}) and (\ref{asmodel}) become
\begin{equation}\label{e0further2}
    \left\{ \begin{array}{l}
    \vec{\Theta}(t) \approx D^{-1} M \big( \vec{P}_{\phi} + \vec{Q}_{\phi} n t \big) + \vec{\lambda} t + \vec{\mu} \\
    \dot{z} \approx Q_{\phi,z} - P_{\phi,z} n t + \gamma
    \end {array} \right.
\end{equation}
where $\vec{P}_{\phi}$ and $\vec{Q}_{\phi}$ are just the functions presented in Eq. (\ref{thieleinnes}) with the angle $\omega + n t_{0}$ replaced with $\phi$, and matrix $M = M_{3 \times 2} = \textrm{diag} (1, 1)$.

Clearly, Eq. (\ref{unchanged}) holds if $\delta \vec{P}_{\phi} = -\delta \vec{\mu}$ and $n \delta \vec{Q}_{\phi} + \vec{Q}_{\phi} \delta n = - \delta \vec{\lambda}$. This means that it is possible to change the values of the components of vector $\vec{P}_{\phi}$ by any amount and a corresponding negative change in the components of vector $\vec{\mu}$ cancels this change exactly. As a result, the components of these vectors can correlate negatively. Also, the components of $\vec{Q}_{\phi}$ can correlate similarly with the components of $\vec{\lambda}$. For RV, $Q_{\phi,z}$ can correlate with $\gamma$, but the product $P_{\phi,z} n$ has no corresponding parameter to correlate with. These are just the correlations described by Eisner \& Kulkarni (2001a, 2001b, 2002).

From Eq. (\ref{e0further2}) it is also clear that the orbital frequency can correlate with $P_{\phi,z}$ and $\vec{Q}_{\phi}$ making the detection of planetary signal harder and broadening the densities of the corresponding parameters.

\begin{figure}
\begin{center}
\includegraphics[angle=270, width=4.0cm, totalheight=3.0cm]{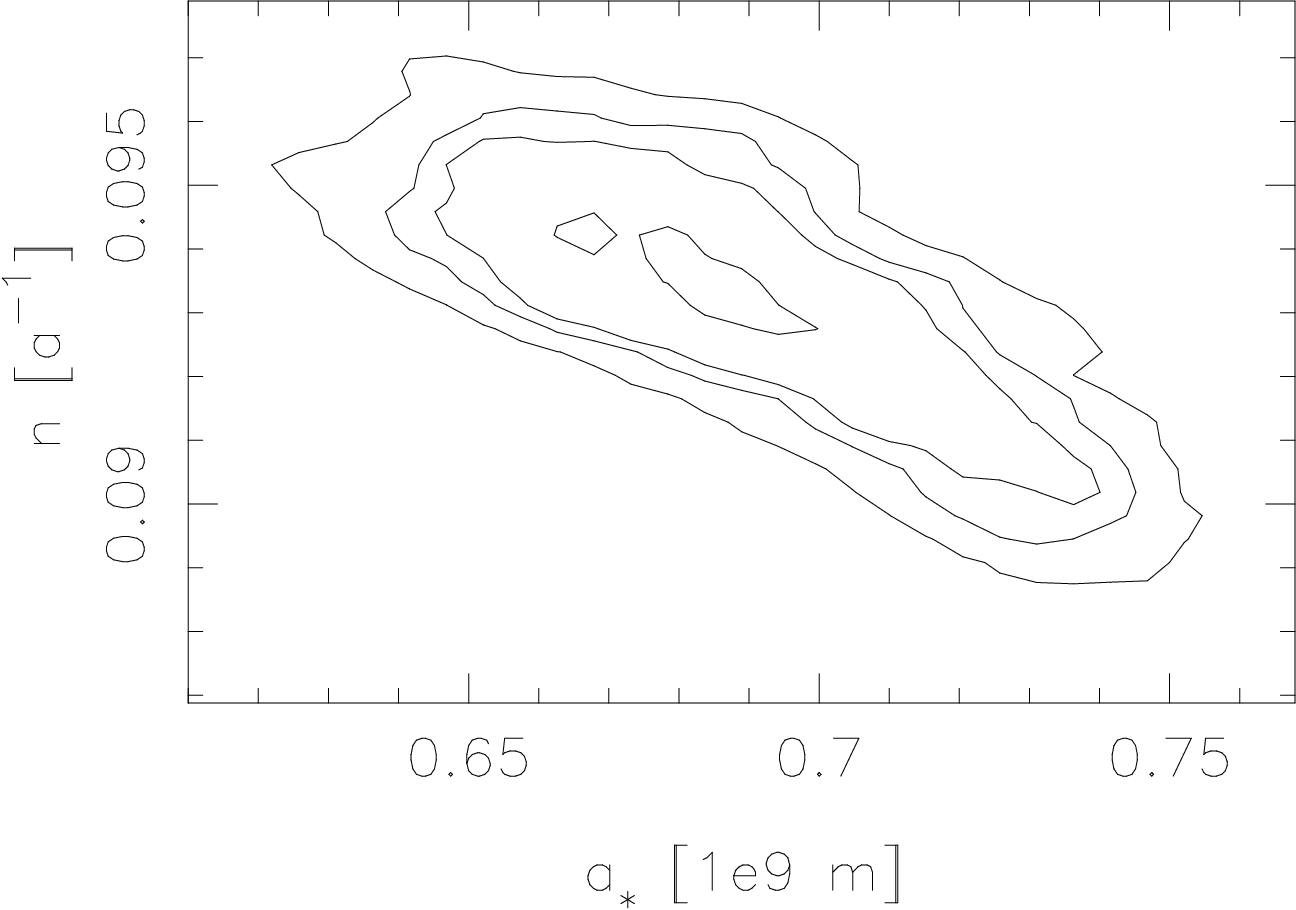}
\includegraphics[angle=270, width=4.0cm, totalheight=3.0cm]{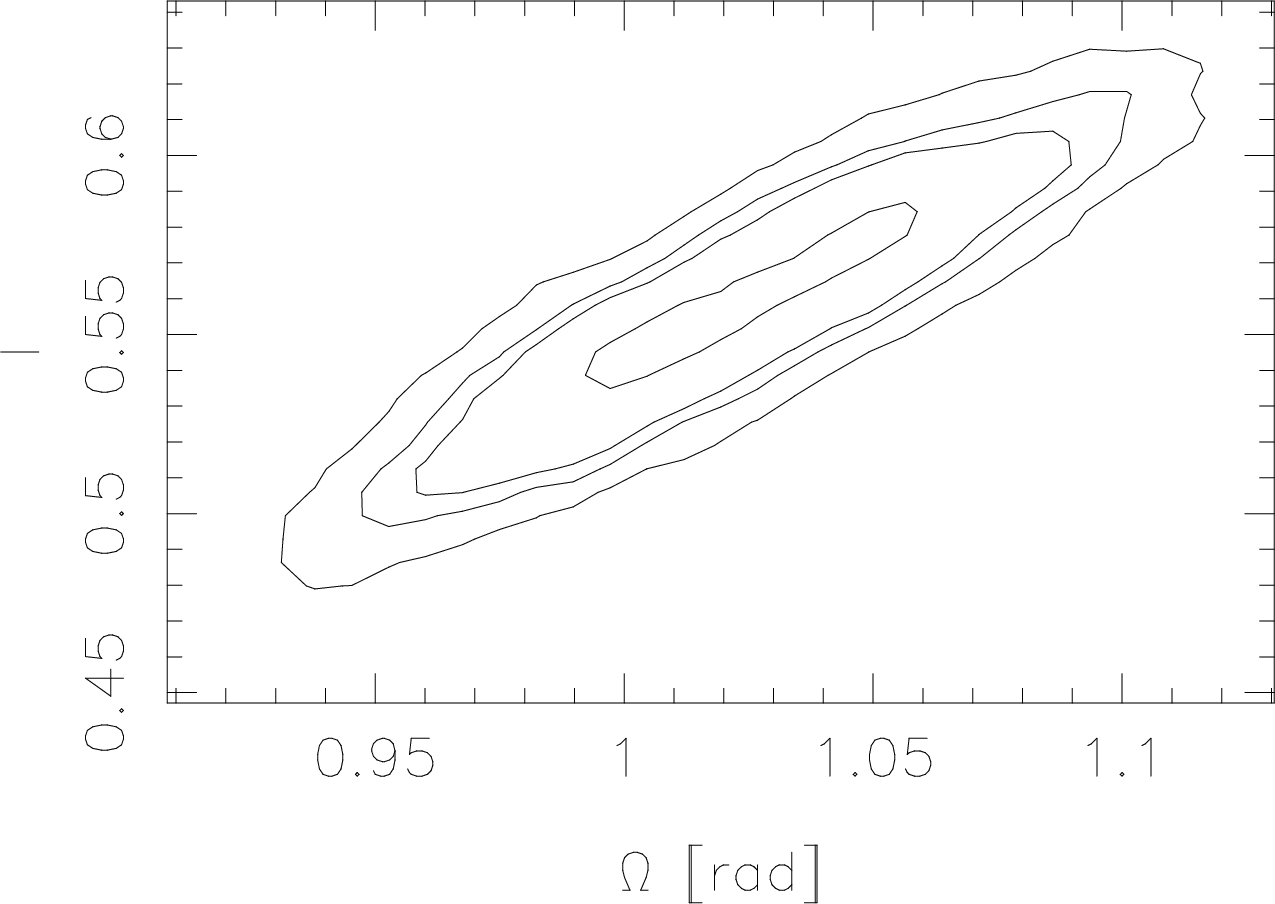}
\end{center}
\caption{Equiprobability contours containing 99\%, 95\%, 90\%, and 50\% of parameter PDF's showing the densities of and correlations between parameters $a_{\star}$ and $n$ and parameters $I$ and $\Omega$. The simulated system has a long-period planet with $T_{A}/P = T_{RV}/P \approx 1/4$ ($T_{RV} = T_{A} = 3.0$ years).} \label{long period detection}
\end{figure}

Despite the existence of and  due to the partially complementary nature of these correlations, it is possible to detect the periodic signal of a long-period planetary companion in the Bayesian model selection sense. The equiprobability contours of parameter combinations $(n, a_{\star})$ and $(I, \Omega)$ with the simulated system S1 are shown in Fig. \ref{long period detection} for $T_{A} = T_{RV} =$ 3.0 years, which is approximately one fourth of the orbital period. The Bayesian model probabilities were found to satisfy the condition of Eq. \ref{best_model}, and the contours in Fig. \ref{long period detection} demonstrate that it is indeed possible to detect extrasolar planetary companions even if the observational timeline is shorter than the orbital period. Also, since clearly $I < 1$, the planetary nature of these companions can be verified in this scenario.

When using the two data sources, the planetary signal could not be detected for $T_{A} = T_{RV} < 3.0$ years. For astrometric or RV measurements alone, this signal was found to be undetectable for $T_{A} = T_{RV} < 10.0$ years, which clearly demonstrates the advantages of the Bayesian inference of multiple datasets.

\subsection{Astrometric snapshots and detection thresholds}

Astrometric observations with the property $T_{A} < P < T_{RV}$ are called astrometric snapshots. This definition is made because the astrometric observations are now made in a fraction of the time interval of the RV observations incapable of separating $m$ and $\sin i$. We modified the simulated system S1 by fixing $T_{RV} = 20.0$ years and denoting this by S2.

\begin{figure}
\begin{center}
\includegraphics[angle=270, width=4.0cm, totalheight=3.0cm]{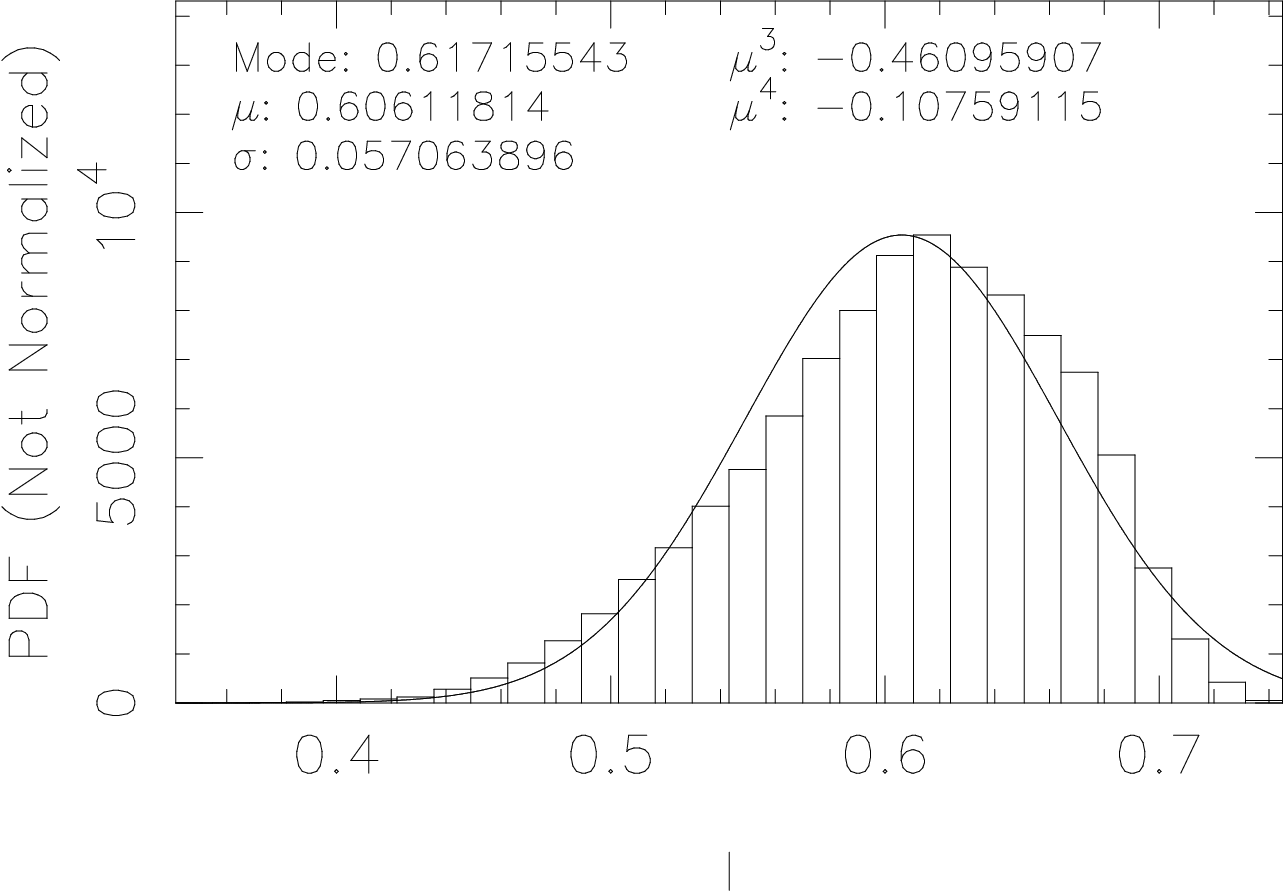}
\includegraphics[angle=270, width=4.0cm, totalheight=3.0cm]{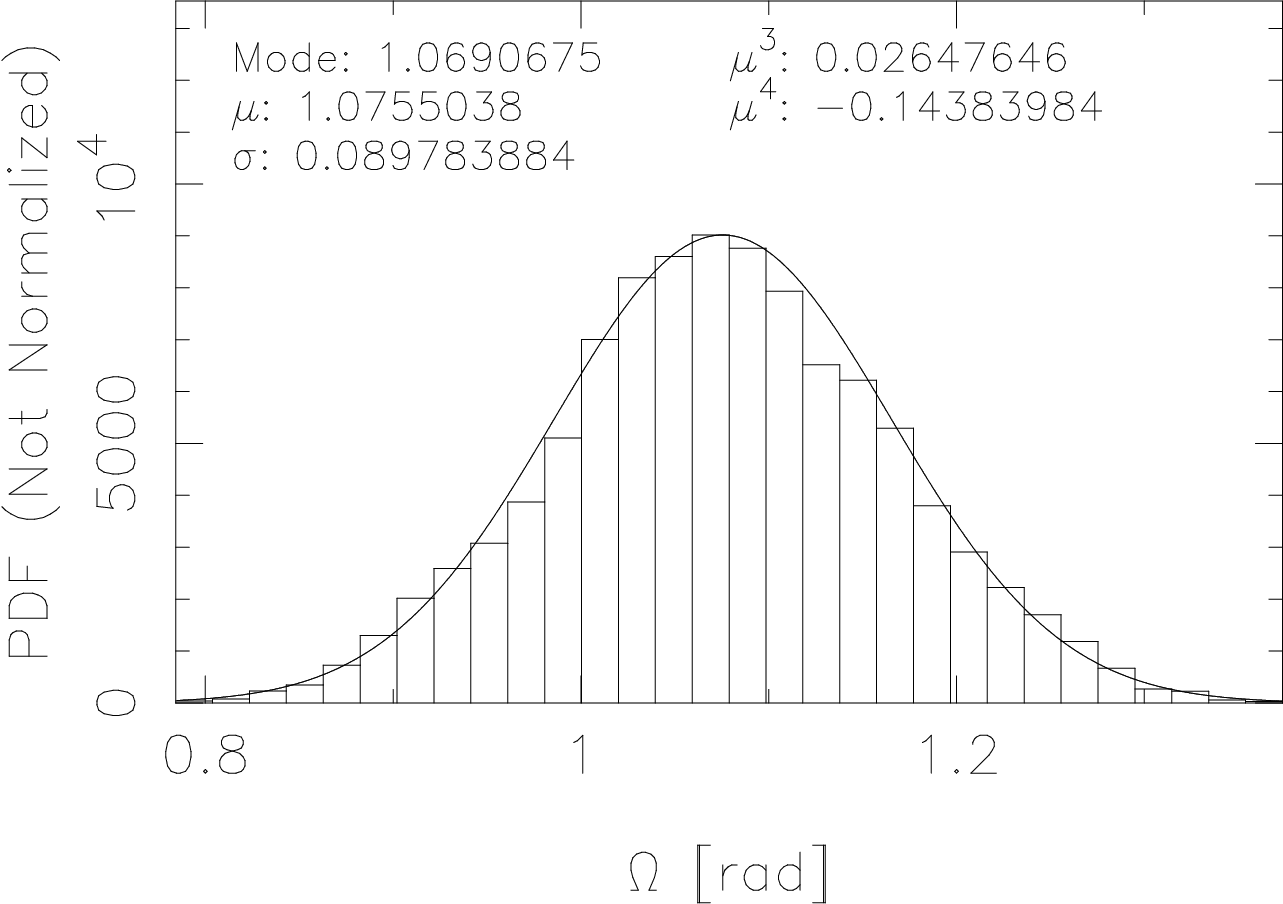}
\end{center}
\caption{PDF's of parameters $I$ and $\Omega$ in a snapshot scenario with $T_{A}/P \approx 1/11$ ($T_{A} = 1.0$ years). The mode, mean ($\mu$), standard deviation ($\sigma$), skewness ($\mu^{3}$), and kurtosis ($\mu^{4}$) of the densities are shown. The solid curve is a Gaussian function $N(\mu , \sigma^{2})$.} \label{angle densities}
\end{figure}

\begin{figure}
\begin{center}
\includegraphics[angle=270, width=4.0cm, totalheight=3.0cm]{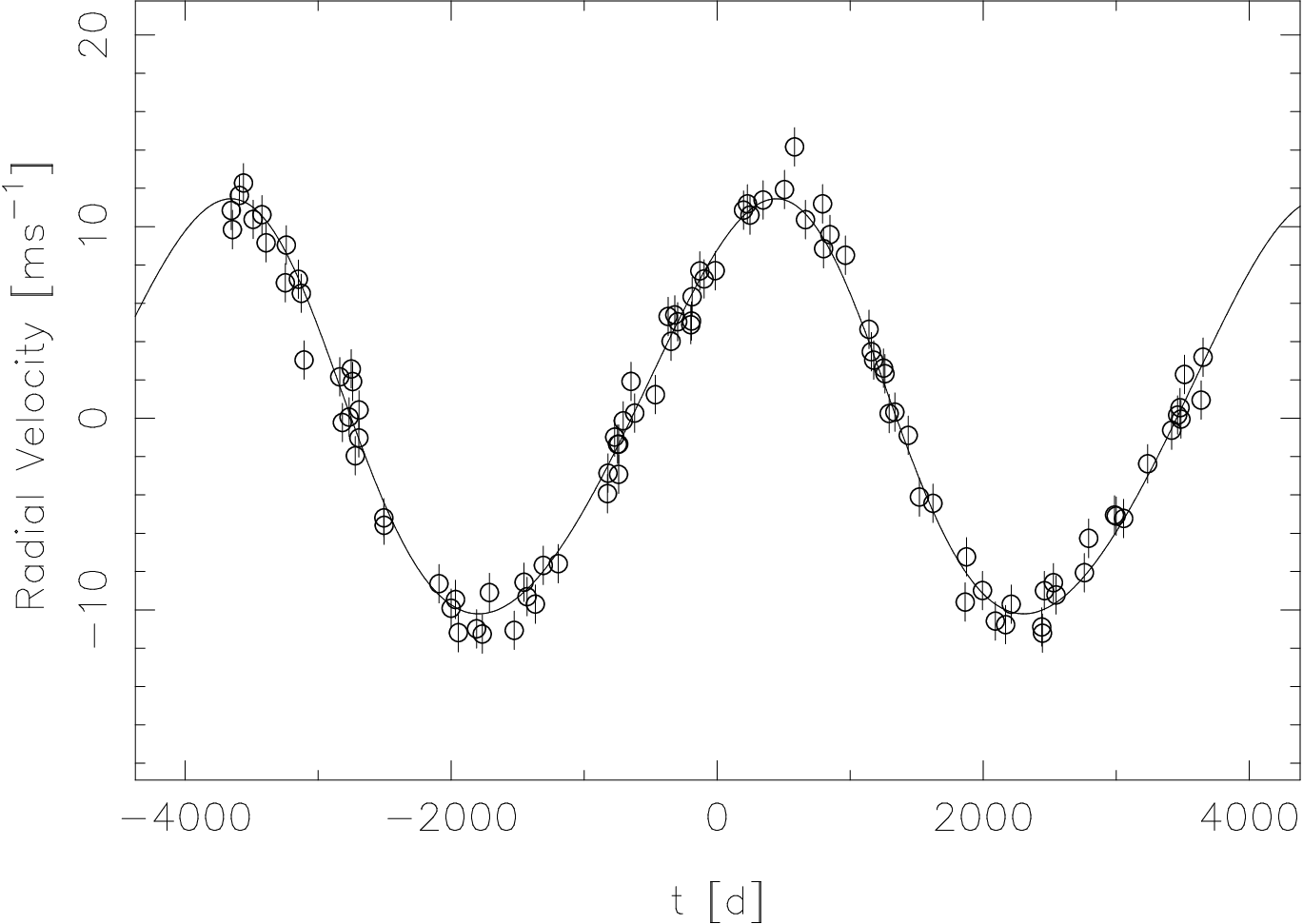}
\includegraphics[angle=270, width=4.0cm, totalheight=3.0cm]{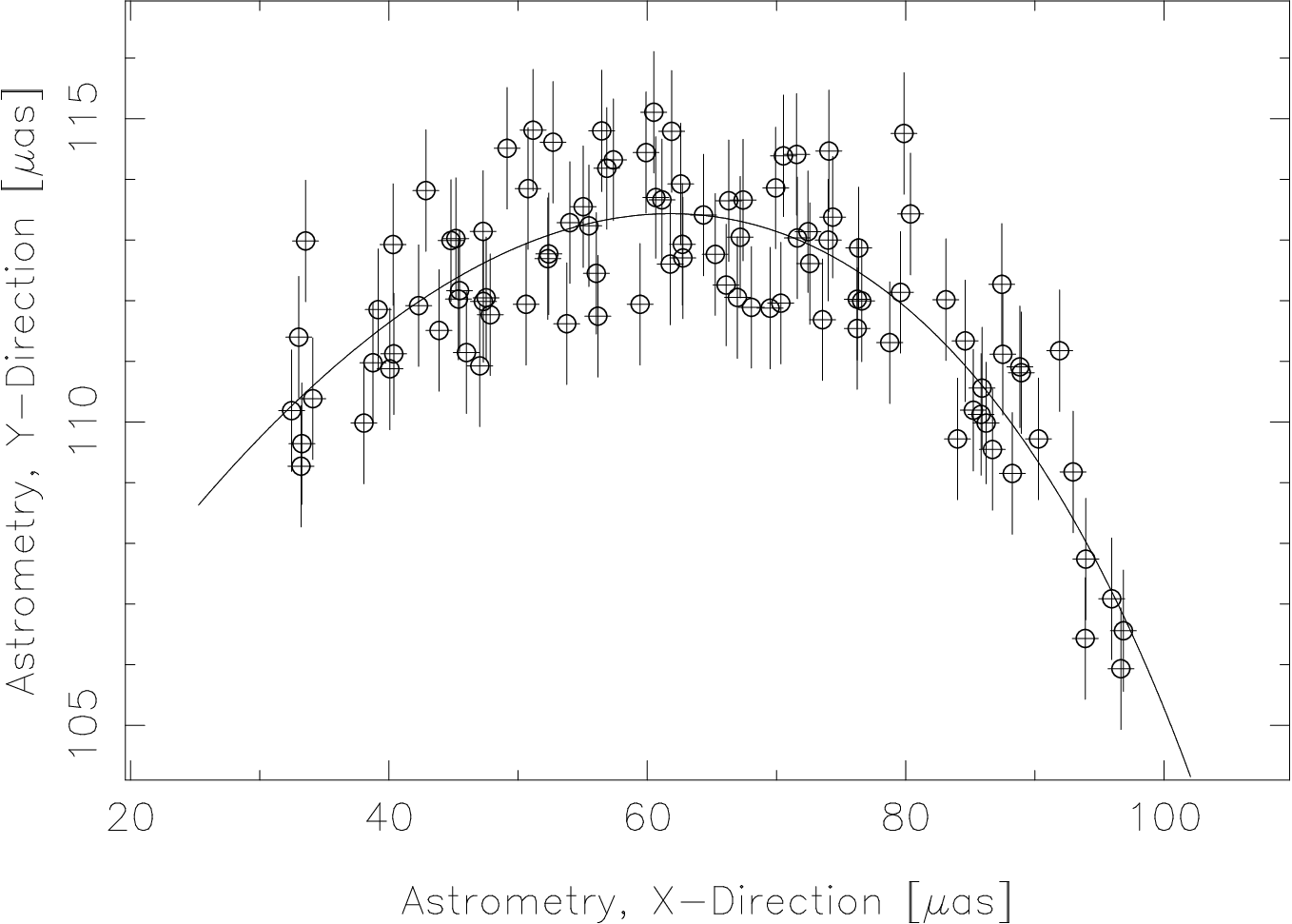}
\end{center}
\caption{Simulated RV and astrometric measurements and the maximum \emph{a posteriori} orbit.} \label{orbit_fig}
\end{figure}

Using Bayesian inference between RV and astrometric measurements made it possible to fit all the parameters in the model, including $I$ and $\Omega$, to the measurements, even when the signal of the planetary companion could not be detected using astrometric observations alone. For the simulated system S2, the condition in Eq. \ref{best_model} was satisfied for values of $T_{A}$ as low as 1.0 years, which is less than one tenth of the orbital period. The corresponding densities of $I$ and $\Omega$ in this scenario are shown in Fig. \ref{angle densities}. It can be seen that the maximum \emph{a posteriori} estimates of these densities are very close to the values selected for the simulated system. This implies that the true mass of the stellar companion is obtainable. Also, the density of $I$ shows that certainly $I < 1$, implying that a companion of planetary mass has indeed been detected as claimed. The simulated RV and astrometric measurements and the maximum \emph{a posteriori} orbit are shown in Fig. \ref{orbit_fig} for this snapshot scenario.

\begin{figure}
\begin{center}
\includegraphics[angle=270, width=4.0cm, totalheight=3.0cm]{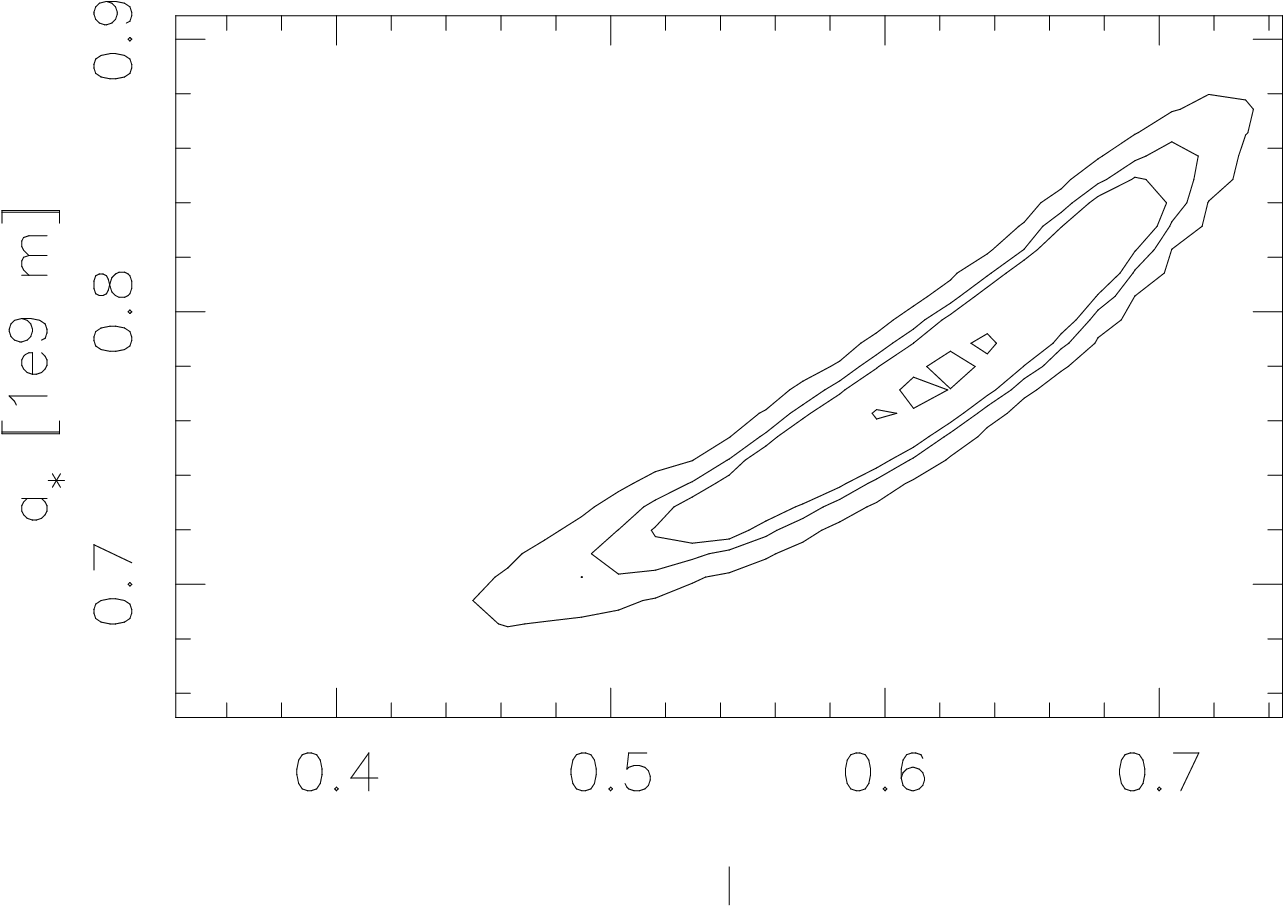}
\includegraphics[angle=270, width=4.0cm, totalheight=3.0cm]{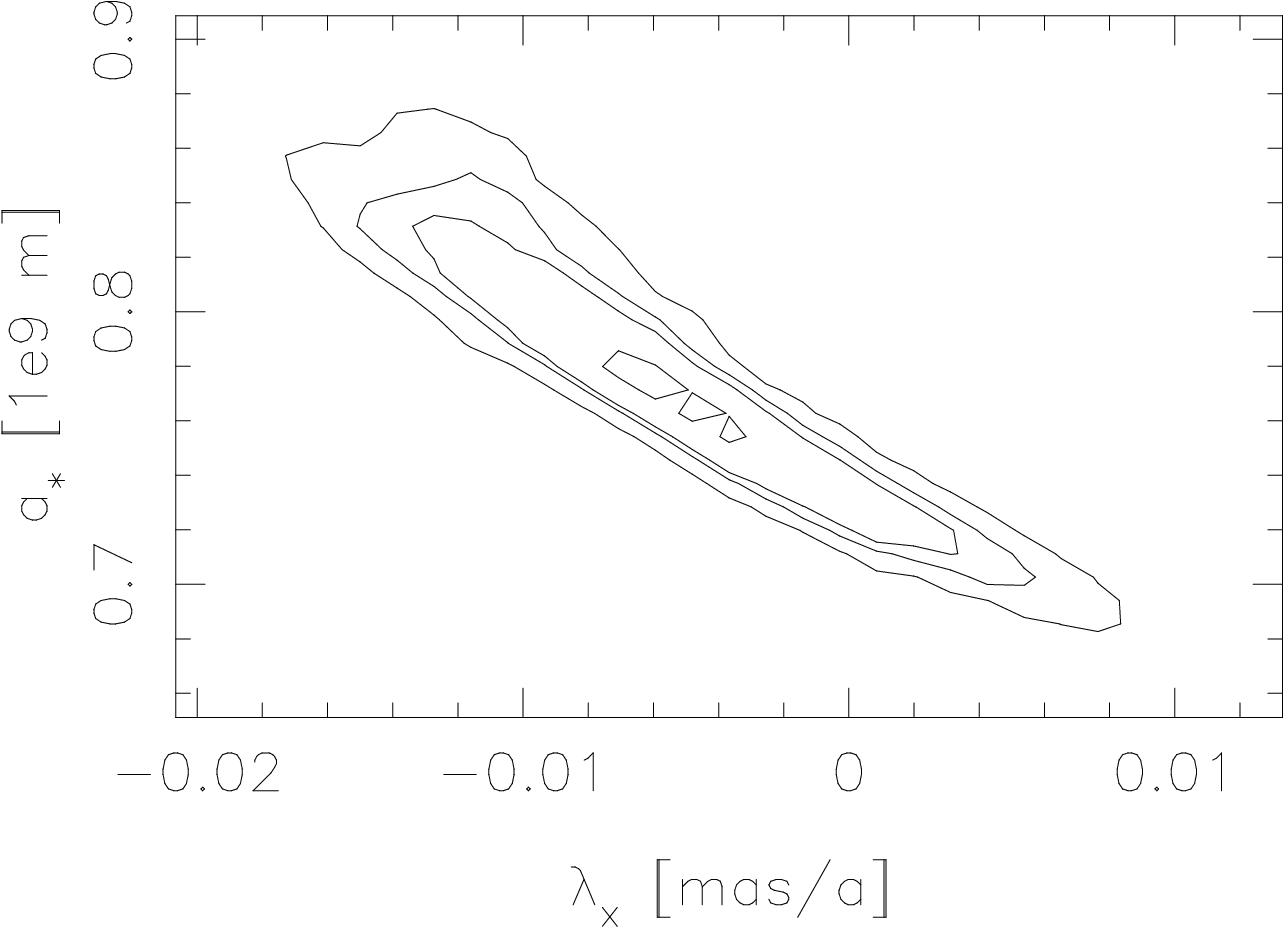}
\end{center}
\caption{Equiprobability contours containing 99\%, 95\%, 90\%, and 50\% of parameter probability densities. The correlations between $a_{\star}$ and $I$ and of $a_{\star}$ and $\lambda_{x}$ in a snapshot scenario with $T_{A}/P \approx 1/11$ ($T_{A} = 1.0$ years).} \label{inclination correlations}
\end{figure}

The reason the parameters $I$ and $\Omega$ can be fitted is that the parameters in the RV model are now well-constrained by the RV measurements. The only possibility for Eq. (\ref{unchanged}) to be true is that $a_{\star} n \sin i$, the amplitude of RV variations, remains unaltered even if $a_{\star}$ and $i$ do not. This implies that $a_{\star} \propto (1-I^{2})^{-1/2}$, which is the correlation observed in the parameter densities of $a_{\star}$ and $I$ (Fig. \ref{inclination correlations}). Because of this correlation, the reference frame parameters of astrometry can correlate freely with $a_{\star}$. This is also demonstrated in Fig. \ref{inclination correlations}.

\section{Case study: HD 154345b}

\begin{figure}
\begin{center}
\includegraphics[angle=270, width=8.0cm, totalheight=6.0cm]{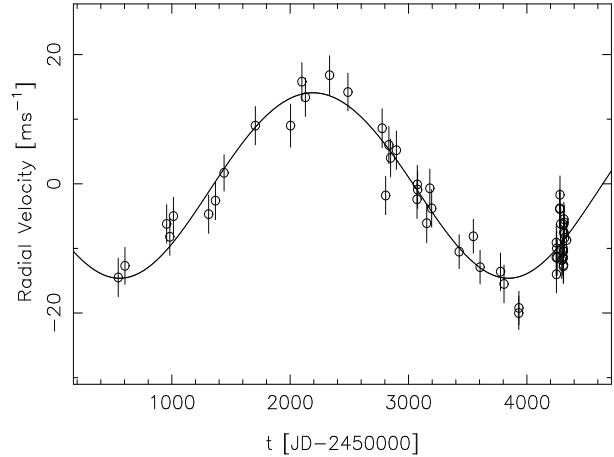}
\end{center}
\caption{RV measurements of HD 154345 and the maximum \emph{a posteriori} orbit of the planetary companion.} \label{HD154345_curve}
\end{figure}

Recently, Wright et al. (2008, hereafter W08) reported a detection of a Jupiter analog orbiting a G8 dwarf HD 154345. They claim that astrometric measurements over its 9-year period would determine the orientation of the orbital plane and as a consequence the true mass. The data published in W08 was re-examined and the orbital solution found using MCMC. These data have 55 measurements over a period of 10.4 years. The largest gap between two subsequent observations within these measurements is 352 days. The orbital parameters were calculated assuming the same jitter level as in W08, 2.5ms$^{-1}$, and are listed in Table \ref{HD 154345 orbit}. This Table shows the MAP estimates of the parameters and their 99\% confidence sets (CS). Missing confidence sets indicate that the posterior density of the corresponding parameter has significant values everywhere in its parameter space. The results of Wright et al. (2008) with 99\% confidence limits are shown for comparison. The corresponding fit is shown in Fig. \ref{HD154345_curve}. The large uncertainties of parameters $\omega$ and $t_{0}$ (their 99\% Bayesian confidence sets are equal to their parameter spaces) stem from these parameters being meaningless for circular orbits.

Regardless of the fact that in W08 the RV movement of the host star had been subtracted from the data, the constant movement parameter $\gamma$ was fitted as well to be able to take its uncertainty and its effect on the orbital parameters into account. The orbital parameters were found consistent with but their confidence intervals smaller than those reported in W08. This difference in the confidence intervals takes place likely because the Bootstrap method was used to assess the parameter errors in W08 instead of direct sampling of the posterior density.

\begin{table}
\center
\caption{The solution and error estimates of the HD 154345 system.\label{HD 154345 orbit}}
\begin{tabular}{l|c|c|c}
  \hline \hline
    Parameter & MAP & 99\% CS & Wright et al. \\
  \hline
    $P$ [years] & 9.06 & [8.56, 9.72] & 9.15 $\pm$ 0.67 \\
    $e$ & 0.02 & [0.00, 0.17] & 0.044 $\pm$ 0.118 \\
    $\omega$ [$^{\circ}$] & 90 & - & 68 \\
    $t_{0} [JD]$ & 245000 & - & 2452830 $\pm$ 850 \\
    $m_{p} \sin i$ [M$_{\textrm{Jup}}$] & 0.95 & [0.80, 1.10] & 0.947 $\pm$ 0.232 \\
    $a$ [AU] & 4.16 & [4.01, 4.37] & 4.19 $\pm$ 0.67 \\
    $\gamma$ [ms$^{-1}$] & 0.01 & [-1.42, 1.42] & 0 \\
  \hline \hline
\end{tabular}
\end{table}

Astrometric measurements with $\sigma_{A} = 1 \mu$as, $N_{A} = 100$, and $T_{A} = $10.0, 5.0, 3.0, 2.0, 1.5, 1.2, 1.0, and 0.8 years were generated to study the detectability of parameters $I$ and $\Omega$. These measurements were generated assuming that there is a planetary companion with orbital parameters in Table \ref{HD 154345 orbit}, and $(I, \Omega) = (0.0, 1.0$ rad). This simulated data was used together with the real RV measurements published in W08 to find the limiting $T_{A}$ for which the true mass of the planet could still be measured with the SIM telescope. The selection $I=0$ was made because changing this value would result in a higher planetary mass and hence in a stronger astrometric signal, making the detection of orbital plane parameters even easier. Regardless of large error bars, we found that it is possible to detect the orbital plane parameters with $T_{A} = 1.0$ years. With this short timeline, the 99\% error bars of the parameters were [-0.30, 0.22] and [0.39, 1.54] for $I$ and $\Omega$, respectively, demonstrating that it was indeed possible to determine their values.

\section{Conclusions and discussion}

The time needed to make a positive detection of an extrasolar planetary conpanion candidate depends essentially on its orbital period. It is commonly assumed that, to be able to detect the signature of such companion, an observational timeline longer than the orbital period is required. Also, since most of the exoplanet candidates have been detected using the RV method, only the lower limit of their mass is available. With the aid of future space telescopes and accurate astrometric measurements, it will be possible to detect the inclination and thus the true mass of planetary candidates.

We have shown that when high-precision RV and accurate astrometric measurements are both available, it is possible to detect the true mass of stellar companions with observational timelines considerably shorter than their orbital periods. Also, when the RV measurements have a long time span, astrometric measurements can reveal the true mass of a stellar companion in less time than one tenth of the orbital period of the system. This ability is also demonstrated using the RV measurements of HD154345 as an example. We find that, having these measurements with $T_{RV} = 10.4$ years in hand, astrometric observations with SIM telescope are sufficient for obtaining the true mass, within a single year.

Bayesian inference plays an important role when extracting information from several sources of measurements. The ability to use RV and astrometric measurements simultaneously makes it possible to employ observational timelines below the orbital ones and still be able to make positive exoplanet detections, thus helping to extract the maximum amount of information from measurements and increasing the time efficiency of observations.

In a forthcoming study, we plan to study the inclusion of additional transit-photometry measurements to further tighten the parameter probability densities in transiting scenarios. Also, the approach used here should be extended to systems with two or more planetary companions.

\begin{acknowledgements}
S. Kotiranta was supported by the Jenny \& Antti Wihuri foundation, and M. Kaasalainen by the Academy of Finland (project ``New mathematical methods in planetary and galactic research''). We would like to thank Dr. J. Wright and the two anonymous referees for pointing out errors and inaccuracies in the manuscript.
\end{acknowledgements}


\end{document}